# Radio frequency spectrum analyzer with a 5 THz bandwidth based on nonlinear optics in a CMOS-compatible high-index doped silica waveguide

Yuhua Li,[1,2] Zhe Kang,[3,4] Kun Zhu,[2] Shiqi Ai,[2] Xiang Wang,[5] Roy R. Davidson,[5] Yan Wu,[1] Roberto Morandotti,[6] Brent E. Little,[7] David J. Moss,[8] and Sai Tak Chu[2]

[1]Key Laboratory of Optical Field Manipulation of Zhejiang Province, Department of Physics, Zhejiang Sci-Tech University, Hangzhou, Zhejiang 310018, China
[2]Department of Physics, City University of Hong Kong, Kowloon Tong, Hong Kong 999077, China
[3]Ningbo Research Institute, Zhejiang University, Ningbo 310000, China
[4]Centre for Optical and Electromagnetic Research, National Engineering Research Center for Optical Instruments, Zhejiang University, Hangzhou 315000, China
[5]QXP Technology, Xi'an 710311, China
[6]INRS-Énergie, Matériaux et Télécommunications, Varennes J3X 1S2, Canada
[7]State Key Laboratory of Transient Optics and Photonics, Xi'an Institute of Optics and Precision Mechanics, Chinese Academy of Science, Xi'an 710119, China
[8]Optical Sciences Centre, Swinburne University of Technology, Hawthorn, VIC 3122, Australia

**We report an all-optical radio-frequency (RF) spectrum analyzer with a bandwidth greater than 5 terahertz (THz), based on a 50-cm long spiral waveguide in a CMOS-compatible high-index doped silica platform. By carefully mapping out the dispersion profile of the waveguides for different thicknesses, we identify the optimal design to achieve near zero dispersion in the C-band. To demonstrate the capability of the RF spectrum analyzer, we measure the optical output of a femtosecond fiber laser with an ultrafast optical RF spectrum in the terahertz regime.**

All optical radio frequency (RF) spectral measurements provide an effective way to analyze ultrafast signals, and have been applied to optical performance monitoring of optical telecommunication systems, ultrafast optical signal characterization and microwave photonics [1-3]. Traditional RF spectral measurement techniques are based on photon detection combined with an electrical spectrum analyzer. Conversely, an all-optical RF spectrum analyzer measures the RF frequency spectrum by detecting the optical signal directly in the optical domain, via the nonlinear optical Kerr effect, allowing it to break the electronic bottleneck arising from the optical-to-electrical conversion. The all-optical approach allows the operation bandwidth to extend into the terahertz regime [4-11]. All-optical RF spectrum analyzers have been demonstrated in a number of nonlinear waveguide platforms, including highly nonlinear fiber (HNLF) [4,5], silicon [6,7], chalcogenide glass [8,9], silicon nitride [10], and doped silica glass waveguide [11]. When implemented on the integrated waveguide platforms, the photonic-chip based RF spectrum analyzer (PC-RFSA) has the advantages of being compact and allowing the integration of additional optical structures for enhanced functionality [12-18]. Previously, Ferrera *et. al.* demonstrated a broadband PC-RFSA with a 4-cm long high-index doped glass waveguide and obtained a bandwidth of 2.5 THz in the telecom C-band [11]. Despite the wide bandwidth that it achieved, the waveguide structures used in the work were not optimized for C-band operation and had a zero dispersion wavelength (ZDW) at the upper edge of the band at around 1560 nm, instead of at the middle of the band at 1547.5 nm. Due to the non-ideal ZDW location, the achieved wide bandwidth of 2.5 THz was at the expense of using a shorter waveguide and reduced sensitivity.

Here, we demonstrate a high-index doped silica based PC-RFSA with improved performance by designing a waveguide structure with low dispersion slope and optimized dispersion for the C-band operation having ZDW at near 1547 nm. We obtained a bandwidth of 5 THz that covers the entire C-band using a 50-cm long waveguide. With the optimized waveguide dispersion, one can obtain broadband operation even with a longer waveguide and results in a higher signal sensitivity.

Figure 1 shows the principle of operation of the PC-RFSA. An optical signal under test (SUT) having an RF spectrum described by $S_{RF}(\omega)$ and a temporal intensity power spectrum given by $|\int I(t)exp(i\omega t)dt|^2$ is mixed with a continuous-wave (CW) probe beam, with electric field given by $E_0 exp(-i\omega_0 t)$ ), and both are then launched into a nonlinear medium [4]. Due to the near-instantaneous response of the cross-phase modulation (XPM) effect [19] in the medium, the phase of the CW probe beam is modulated by the intensity of the SUT, with a modulation index of $2\gamma L_{eff}$, where $\gamma$ and $L_{eff}$ are the nonlinear parameter and effective propagation length of the waveguide, respectively. As the combined signals propagate through the nonlinear waveguide, frequency modulation sidebands are generated around $\omega_0$, with their combined optical spectrum being proportional to $(E_0 \gamma L_{eff})^2 S_{RF}(\omega - \omega_0)$ due to the XPM [4]. This results in the RF spectrum $S_{RF}(\omega)$ of the SUT being mapped onto the sideband of the optical spectrum of the probe light, which can then be measured directly by an optical spectrum analyzer (OSA).

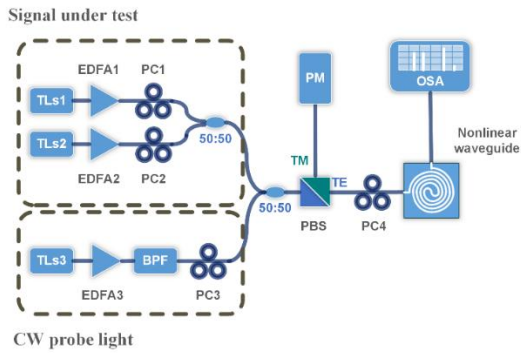

Fig. 1. Experimental setup for the PC-RFSA. (TLs: tunable laser source; EDFA: erbium-doped fiber amplifier; PC: polarization controller; BPF: bandpass filter; PBS: polarization beam splitter; PM: power meter.)

Since this is an all-optical operation, in principle it can achieve unlimited bandwidth as it is not limited by the electronic bandwidth from optical-to-electrical conversion. In practice, effects such as group velocity mismatch between the SUT and the CW probe light, limit the bandwidth according to

$$\Delta f_{max} = \frac{1}{L_{eff} \left| \Delta\lambda \left( 2D + \frac{\partial D}{\partial \lambda} \Delta\lambda \right) \right|} \quad (1)$$

where $D$ is the dispersion of the medium, $\partial D/\partial \lambda$ is the dispersion slope at the wavelength where the SUT is centered, and $\Delta\lambda$ is the wavelength spacing between the SUT and the CW probe light [11]. Therefore, waveguides with both low dispersion and low dispersion slope are crucial to maximize the device bandwidth. The optimal bandwidth for a given probe to SUT wavelength separation of $\Delta\lambda$ is obtained by setting the probe and SUT wavelength symmetrically with respect to the ZDW [4]. For C-band application, the ideal ZDW location should be at the center of the band of around 1547 nm.

To optimize the waveguide for the PC-RFSA for operation in the C-band, we calculate the waveguide dispersion as a function of geometry and wavelength. Figures 2(a) and 2(c) show the TE and TM dispersion maps as a function of waveguide thickness for a width of 2 µm. The wavelength dependent material refractive indices used in the calculation are obtained from film index measurements to account for material dispersion. The gray dashed lines in the figures reveal the zero-dispersion points over the mapped space and they indicate that for operation with zero-dispersion wavelength of around 1547 nm, the thicknesses of the waveguide is 1.27 µm for TE and 1.35 µm for TM modes. Figures 2(b) and 2(d) show the calculated nonlinear parameter $\gamma$ for the TE and TM modes of the waveguide versus thickness. The nonlinearity increases with decreasing thickness and wavelength due to stronger optical confinement. However, at 1547 nm there is minimal change in $\gamma$ for thicknesses from 0.5 µm to 1.5 µm.

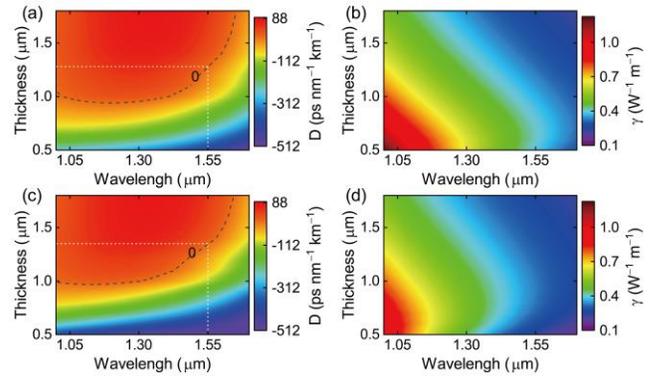

Fig. 2. (a) Calculated dispersion and (b) nonlinearity for TE mode; and (c) calculated dispersion and (d) nonlinearity for TM mode, for waveguides with 2 µm width and variable thicknesses.

A series of 50-cm and 100-cm long spiral waveguides with a width of 2 µm and thickness ranging from 1.00 µm to 1.50 µm in steps of 0.25 µm were fabricated using the process described in [20-26] to investigate their performance as PC-RFSAs. The inset in Fig. 3 shows a scanning electron microscope (SEM) image of the fabricated buried waveguide with cross-section 2 µm × 1.25 µm. The core refractive index is 1.7 and the cladding index is 1.45. It can be seen that there is a sidewall angle of approximately 8° resulting from the etching process. With the mode transformers at the input and output ends of the spirals, the coupling loss between the optical fiber and waveguide is about 1 dB/facet, while the propagation loss varies from 0.06 to 0.1 dB/cm at 1547 nm for the different thicknesses. The calculated dispersion curve for a 2 µm × 1.25 µm waveguide (Fig. 3) shows that the ZDWs are at 1542 nm and 1503 nm for the TE and TM modes, respectively. At 1547 nm, the desired operation point for C-band application, the dispersion is −2.6 ps·nm$^{-1}$·km$^{-1}$ for the TE mode and −22.1 ps·nm$^{-1}$·km$^{-1}$ for the TM mode, while the dispersion slopes are −0.47 ps·nm$^{-2}$·km$^{-1}$ and −0.55 ps·nm$^{-2}$·km$^{-1}$ for TE and TM modes, respectively. The dispersion for the TE mode across the whole C and L bands remains low at < 50 ps·nm$^{-1}$·km$^{-1}$.

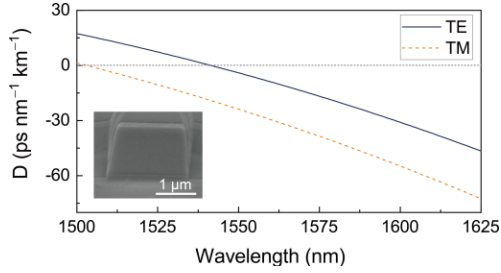

Fig. 3. Simulated dispersion of the fabricated waveguide with 1.25 μm thickness. Inset: SEM image of the cross-section geometry of the waveguide.

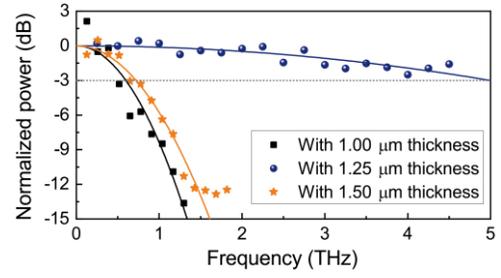

Fig. 4. Normalized XPM power as a function of the frequency spacing between TLs1 and TLs2 for TE mode. The measured data are fit with the 2nd order polynomial function.

The setup for the PC-RFSA measurement is shown in Fig. 1 where tunable narrow-linewidth laser sources, TLs1 and TLs2, with wavelengths at $\lambda_1$ and $\lambda_2$, respectively, served as the SUT and TLs3 with wavelength $\lambda_3$ as the CW probe. The two SUT signals were initially set at 1547 nm to fully utilize the entire C-band spectrum. Their spacings were then detuned in opposite directions from the C-band center to simulate the broadening of the SUT frequency spectrum in the measurement. The probe signal from TLs3 was fixed at 1601 nm, far away from the SUT signals to avoid spectral overlap between the four-wave-mixing signal from the SUT and XPM signal from the probe, when the bandwidth measurement is performed in the whole C-band. PCs, PBS and PM were used to ensure that the polarization states of all the three signals were aligned to the TE or TM mode of the waveguide.

**Table 1 Measured bandwidth for different platforms.**

| Platform | Waveguide length (cm) | Bandwidth (THz) | Ref. |
|---|---|---|---|
| HNLF | $6 \times 10^4$ | 0.8 | [4] |
| SiN | 2 | 0.9 | [10] |
| As$_2$S$_3$ | 16 | 3.2 | [8] |
| Silicon | 1.5 | 1.6 | [6] |
| High-index doped silica | 4 | 2.5 | [11] |
| High-index doped silica | 50 | 5.0 | This work |

Figure 4 shows the normalized XPM power as a function of frequency separation between TLs1 and TLs2 for the set of 50-cm spiral waveguides having thicknesses between 1.00 μm to 1.50 μm for the TE mode with each of the measured data fitted to parabolic curves. For the 1.25 μm thick waveguide, we obtain a 3-dB bandwidth for the TE mode of ∼ 5 THz and 3.43 THz for the 50 cm and 100 cm waveguides, respectively. Table 1 summarizes the results from the literature, showing that the 5 THz bandwidth obtained in this work corresponds to the highest value reported so far. We also obtained a XPM walk-off delay, defined as $2\gamma P_0 T_0 / |\Delta\lambda(2D - \partial D/\partial\lambda \Delta\lambda)|$ [8] of 70 for the 1.25 μm thick waveguide, indicating the walk-off delay influence on XPM is extremely small and the nonlinear interaction is maintained over large $\Delta\lambda$.

Table 2 lists the measured 3-dB bandwidths for all three fabricated waveguides using the experimental setup in Fig. 1 and the same wavelengths. Since the ZDWs are located outside of the C-band for the 1.00 μm and 1.50 μm waveguides, their obtained bandwidths are limited. However, one can shorten the waveguide to further increase the bandwidth if desired, similar to [11].

**Table 2 Measured bandwidth of waveguides with different thicknesses for TE mode.**

| Waveguide thickness (μm) | Simulated dispersion at 1547 nm (ps nm$^{-1}$ km$^{-1}$) | Dispersion slope at 1547 nm (ps nm$^{-2}$ km$^{-1}$) | Bandwidth (THz) |
|---|---|---|---|
| 1.00 | −54.9 | −0.54 | 0.59 |
| 1.25 | −2.6 | −0.47 | 5.00 |
| 1.50 | 36.1 | −0.41 | 0.73 |

Figure 5 compares the PC-RFSA sensitivity for a 50-cm and a 100-cm spiral waveguide, with the sensitivity defined as the power level of the generated XPM signal that is 3 dB above the OSA (Yokogawa, AQ6370D) noise floor [2] for a given probe power. This represents the minimum power level that the PC-RFSA can detect the 3-dB bandwidth of the SUT. The frequency separation between the SUT signals was set to 625 GHz, with $\lambda_1$, $\lambda_2$, $\lambda_3$ set at 1550 nm, 1555 nm, 1580 nm, respectively. By setting the probe power to 0 dBm, the relationship between the input SUT power and generated XPM power from the 50-cm and 100-cm long spirals is shown in Fig. 5. Due to the accumulated nonlinear effect and low propagation loss of the waveguide in the C-band, the 100-cm long spiral waveguide produced a higher XPM signal. At the lowest noise floor setting of the OSA at −88 dBm, the minimum resolvable input SUT power value was 1.5 dBm for the 100-cm long waveguide, which was 2 dB lower than that for the 50-cm long waveguide.

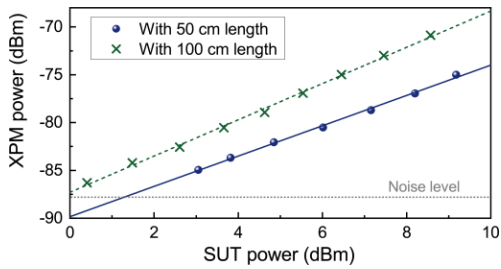

Fig. 5. The XPM sideband power as a function of the averaged input SUT power for the 1.25 μm thick waveguides and different lengths (TE mode is considered and the SUT frequency is fixed on 625 GHz).

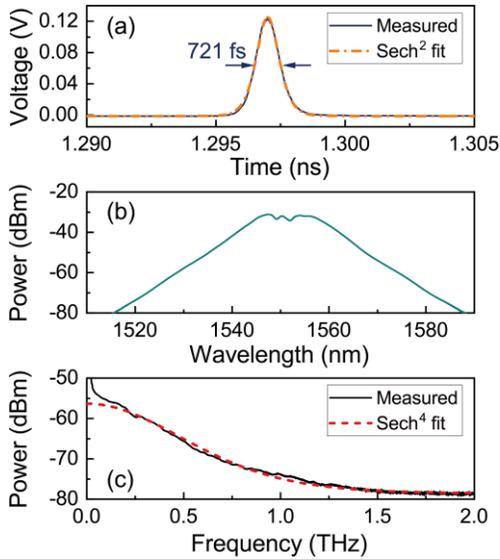

Fig. 6. (a) Measured and fit pulse shape with envelope width of 721 fs. (b) Optical spectrum of femtosecond laser recorded by OSA. (c) (solid curve) RF spectrum recorded by PC-RFSA, (dotted curve) sech$^4$ fit curve.

To investigate the distortion of the PC-RFSA, we measured the RF spectrum of a femtosecond fiber laser (PriTel, with a pulse width of 721 fs and repetition rate of 20 MHz), centered at 1547 nm with an average power of 2.5 mW. The probe was centered at 1601 nm with power of 45 mW. Figures 6(a) and 6(b) show the temporal intensity from autocorrelation measurements with its corresponding optical spectrum. The RF spectrum of the femtosecond pulse, measured by the PC-RFSA, is depicted in Fig. 6(c). The PC-RFSA measured the RF spectrum as broad as 2 THz. The dashed curve in Fig. 6(c) represents the simulated RF spectrum, fit to a sech$^4$ function [9]. As shown in Fig. 6(c), the measured RF spectrum agrees well with theory, which proves that our PC-RFSA is suitable for RF spectrum analysis of signals with bandwidths in the terahertz regime.

In conclusion, we present a PC-RFSA that achieves an RF bandwidth extending well into the terahertz regime, based on a high-index doped silica waveguide. We investigate the dispersion and nonlinear parameter for different waveguide dimensions. By engineering the dispersion, a 3-dB bandwidth of 5 THz is achieved, which is twice as large as that previously reported [11]. The measurement of the RF spectrum of a femtosecond fiber laser demonstrates its usefulness for frequency spectrum analysis of ultra-broadband optical signals involving frequencies in the terahertz range.

**Funding.** The Natural Science Foundation of Zhejiang Province (LY21F050007); Zhejiang Provincial Natural Science Foundation of China (LGG21A040003); The National Natural Science Foundation of China (62075188).

**Disclosures.** The authors declare no conflicts of interest.

†These authors contributed equally to this work.

**REFERENCES**

1. T. D. Vo, M. D. Pelusi, J. Schröder, F. Luan, S. J. Madden, D. Y. Choi, D. A. P. Bulla, B. Luther-Davies, and B. J. Eggleton, Opt. Express **18**, 3938 (2010).
2. I. A. Walmsley and C. Dorrer, Advances in Optics and Photonics **1**, 308 (2009).
3. H. Ou, C. Ye, K. Zhu, Y. Hu, and H. Fu, J. Lightwave Technol. **28**, 2337 (2010).
4. C. Dorrer and D. Maywar, J. Lightwave Technol. **22**, 266 (2004).
5. L. Chen, Y. Duan, H. Zhou, X. Zhou, C. Zhang, and X. Zhang, Opt. Express **25**, 9416 (2017).
6. B. Corcoran, T. D. Vo, M. D. Pelusi, C. Monat, D.-X. Xu, A. Densmore, R. Ma, S. Janz, D. J. Moss, and B. J. Eggleton, Opt. Express **18**, 20190 (2010).
7. M. Ma, R. Adams, and L. R. Chen, J. Lightwave Technol. **35**, 2622 (2017).
8. M. Pelusi, F. Luan, T. D. Vo, M. R. Lamont, S. J. Madden, D. A. Bulla, D.-Y. Choi, B. Luther-Davies, and B. J. Eggleton, Nat. Photonics **3**, 139 (2009).
9. M. Pelusi, T. D. Vo, F. Luan, S. J. Madden, D.-Y. Choi, D. Bulla, B. Luther-Davies, and B. J. Eggleton, Opt. Express **17**, 9314 (2009).
10. M. R. Dizaji, C. J. Krückel, A. Fülöp, P. A. Andrekson, V. Torres-Company, and L. R. Chen, Opt. Express **25**, 12100 (2017).
11. M. Ferrera, C. Reimer, A. Pasquazi, M. Peccianti, M. Clerici, L. Caspani, S. T. Chu, B. E. Little, R. Morandotti, and D. J. Moss, Opt. Express **22**, 21488 (2014).
12. V. R. Almeida, C. A. Barrios, R. R. Panepucci, and M. Lipson, Nature **431**, 1081 (2004).
13. R. Slavík, F. Parmigiani, J. Kakande, C. Lundström, M. Sjödin, P. A. Andrekson, R. Weerasuriya, S. Sygletos, A. D. Ellis, and L. Grüner-Nielsen, Nat. Photonics **4**, 690 (2010).
14. K. Nozaki, A. Shinya, S. Matsuo, Y. Suzaki, T. Segawa, T. Sato, Y. Kawaguchi, R. Takahashi, and M. Notomi, Nat. Photonics **6**, 248 (2012).
15. D. Ballarini, M. De Giorgi, E. Cancellieri, R. Houdré, E. Giacobino, R. Cingolani, A. Bramati, G. Gigli, and D. Sanvitto, Nature communications **4**, 1 (2013).
16. P. Rani, Y. Kalra, and R. Sinha, Optik **126**, 950 (2015).
17. W. Liu, M. Li, R. S. Guzzon, E. J. Norberg, J. S. Parker, M. Lu, L. A. Coldren, and J. Yao, Nat. Photonics **10**, 190 (2016).
18. L. Cong, Y. K. Srivastava, H. Zhang, X. Zhang, J. Han, and R. Singh, Light: Science & Applications **7**, 1 (2018).
19. S. Radic, D. J. Moss, and B. J. Eggleton, I. P. Kaminow, T. Li, and A. E. Willner, eds. (Academic Press, Burlington, 2008), pp. 759.
20. M. Ferrera, L. Razzari, D. Duchesne, R. Morandotti, Z. Yang, M. Liscidini, J. Sipe, S. Chu, B. Little, and D. Moss, Nat. Photonics **2**, 737 (2008).
21. D. J. Moss, R. Morandotti, A. L. Gaeta, and M. Lipson, "New CMOS-compatible platforms based on silicon nitride and Hydex for nonlinear optics," Nat. Photonics **7**, 597-607 (2013).

22. M. Kues, C. Reimer, B. Wetzel, P. Roztocki, B. E. Little, S. T. Chu, D. J. Moss, and R. Morandotti, l., "An ultra-narrow spectral width passively modelocked laser", Nat. Photonics **11**, no. 3, 159-163 (2017). doi:10.1038/nphoton.2016.271
23. A. Pasquazi M. Peccianti, Y. Park, B. E. Little, S. T. Chu, R. Morandotti, J. Azaña, and D. J. Moss., "Sub-picosecond phase-sensitive optical pulse characterization on a chip", Nat. Photonics **5**, no. 10, 618 - 623 (2011). DOI: 10.1038/nphoton.2011.199.
24. L. Razzari, D. Duchesne, M. Ferrera, R.Morandotti, B.E. Little, S. Chu and D..J. Moss, "CMOS-compatible integrated optical hyper-parametric oscillator," Nat. Photonics **4,** (1) 41-45 (2010).
25. H. Bao, A. Cooper, M. Rowley, L. Di Lauro, J. Sebastian T. Gongora, S. T. Chu, B.rent E. Little, G. -L. Oppo, R. Morandotti, D. J. Moss, B. Wetzel, M. Peccianti and A. Pasquazi, "Laser Cavity-Soliton Micro-Combs", Nat. Photonics **13**, no. 6, 384–389 (2019).
26. M.Peccianti, A.Pasquazi, Y.Park, B.E Little, S.T.Chu, D.J Moss, and R.Morandotti, "Demonstration of an ultrafast nonlinear microcavity modelocked laser", Nat. Communications **3,** 765 (2012). doi:10.1038/ncomms1762 (2012).

## Full References

1. T. D. Vo, M. D. Pelusi, J. Schröder, F. Luan, S. J. Madden, D. Y. Choi, D. A. P. Bulla, B. Luther-Davies, and B. J. Eggleton, "Simultaneous multi-impairment monitoring of 640 Gb/s signals using photonic chip based RF spectrum analyzer," Opt. Express **18**, 3938-3945 (2010).
2. I. A. Walmsley and C. Dorrer, "Characterization of ultrashort electromagnetic pulses," Advances in Optics and Photonics **1**, 308-437 (2009).
3. H. Ou, C. Ye, K. Zhu, Y. Hu, and H. Fu, "Millimeter-wave harmonic signal generation and distribution using a tunable single-resonance microwave photonic filter," J. Lightwave Technol. **28**, 2337-2342 (2010).
4. C. Dorrer and D. Maywar, "RF spectrum analysis of optical signals using nonlinear optics," J. Lightwave Technol. **22**, 266 (2004).
5. L. Chen, Y. Duan, H. Zhou, X. Zhou, C. Zhang, and X. Zhang, "Real-time broadband radio frequency spectrum analyzer based on parametric spectro-temporal analyzer (PASTA)," Opt. Express **25**, 9416-9425 (2017).
6. B. Corcoran, T. D. Vo, M. D. Pelusi, C. Monat, D.-X. Xu, A. Densmore, R. Ma, S. Janz, D. J. Moss, and B. J. Eggleton, "Silicon nanowire based radio-frequency spectrum analyzer," Opt. Express **18**, 20190-20200 (2010).
7. M. Ma, R. Adams, and L. R. Chen, "Integrated photonic chip enabled simultaneous multichannel wideband radio frequency spectrum analyzer," J. Lightwave Technol. **35**, 2622-2628 (2017).
8. M. Pelusi, F. Luan, T. D. Vo, M. R. Lamont, S. J. Madden, D. A. Bulla, D.-Y. Choi, B. Luther-Davies, and B. J. Eggleton, "Photonic-chip-based radio-frequency spectrum analyser with terahertz bandwidth," Nat. Photonics **3**, 139 (2009).
9. M. Pelusi, T. D. Vo, F. Luan, S. J. Madden, D.-Y. Choi, D. Bulla, B. Luther-Davies, and B. J. Eggleton, "Terahertz bandwidth RF spectrum analysis of femtosecond pulses using a chalcogenide chip," Opt. Express **17**, 9314-9322 (2009).
10. M. R. Dizaji, C. J. Krückel, A. Fülöp, P. A. Andrekson, V. Torres-Company, and L. R. Chen, "Silicon-rich nitride waveguides for ultra-broadband nonlinear signal processing," Opt. Express **25**, 12100-12108 (2017).
11. M. Ferrera, C. Reimer, A. Pasquazi, M. Peccianti, M. Clerici, L. Caspani, S. T. Chu, B. E. Little, R. Morandotti, and D. J. Moss, "CMOS compatible integrated all-optical radio frequency spectrum analyzer," Opt. Express **22**, 21488-21498 (2014).
12. V. R. Almeida, C. A. Barrios, R. R. Panepucci, and M. Lipson, "All-optical control of light on a silicon chip," Nature **431**, 1081-1084 (2004).
13. R. Slavík, F. Parmigiani, J. Kakande, C. Lundström, M. Sjödin, P. A. Andrekson, R. Weerasuriya, S. Sygletos, A. D. Ellis, and L. Grüner-Nielsen, "All-optical phase and amplitude regenerator for next-generation telecommunications systems," Nat. Photonics **4**, 690-695 (2010).
14. K. Nozaki, A. Shinya, S. Matsuo, Y. Suzaki, T. Segawa, T. Sato, Y. Kawaguchi, R. Takahashi, and M. Notomi, "Ultralow-power all-optical RAM based on nanocavities," Nat. Photonics **6**, 248-252 (2012).
15. D. Ballarini, M. De Giorgi, E. Cancellieri, R. Houdré, E. Giacobino, R. Cingolani, A. Bramati, G. Gigli, and D. Sanvitto, "All-optical polariton transistor," Nature communications **4**, 1-8 (2013).
16. P. Rani, Y. Kalra, and R. Sinha, "Design of all optical logic gates in photonic crystal waveguides," Optik **126**, 950-955 (2015).
17. W. Liu, M. Li, R. S. Guzzon, E. J. Norberg, J. S. Parker, M. Lu, L. A. Coldren, and J. Yao, "A fully reconfigurable photonic integrated signal processor," Nat. Photonics **10**, 190-195 (2016).
18. L. Cong, Y. K. Srivastava, H. Zhang, X. Zhang, J. Han, and R. Singh, "All-optical active THz metasurfaces for ultrafast polarization switching and dynamic beam splitting," Light: Science & Applications **7**, 1-9 (2018).
19. S. Radic, D. J. Moss, and B. J. Eggleton, "20 - Nonlinear optics in communications: From crippling impairment to ultrafast tools," in Optical Fiber Telecommunications V A (Fifth Edition), I. P. Kaminow, T. Li, and A. E. Willner, eds. (Academic Press, Burlington, 2008), pp. 759-828.
20. M. Ferrera, L. Razzari, D. Duchesne, R. Morandotti, Z. Yang, M. Liscidini, J. Sipe, S. Chu, B. Little, and D. Moss, "Low-power continuous-wave nonlinear optics in doped silica glass integrated waveguide structures," Nat. Photonics **2**, 737-740 (2008).
21. D. J. Moss, R. Morandotti, A. L. Gaeta, and M. Lipson, "New CMOS-compatible platforms based on silicon nitride and Hydex for nonlinear optics," Nat. Photonics **7**, 597-607 (2013).
22. M. Kues, C. Reimer, B. Wetzel, P. Roztocki, B. E. Little, S. T. Chu, D. J. Moss, and R. Morandotti, l., "An ultra-narrow spectral width passively modelocked laser", Nat. Photonics **11**, no. 3, 159-163 (2017). doi:10.1038/nphoton.2016.271
23. A. Pasquazi M. Peccianti, Y. Park, B. E. Little, S. T. Chu, R. Morandotti, J. Azaña, and D. J. Moss., "Sub-picosecond phase-sensitive optical pulse characterization on a chip", Nat. Photonics **5**, no. 10, 618 - 623 (2011). DOI: 10.1038/nphoton.2011.199.
24. L. Razzari, D. Duchesne, M. Ferrera, R.Morandotti, B.E. Little, S. Chu and D..J. Moss, "CMOS-compatible integrated optical hyper-parametric oscillator," Nat. Photonics **4,** (1) 41-45 (2010).
25. H. Bao, A. Cooper, M. Rowley, L. Di Lauro, J. Sebastian T. Gongora, S. T. Chu, B.rent E. Little, G. -L. Oppo, R. Morandotti, D. J. Moss, B. Wetzel, M. Peccianti and A. Pasquazi, "Laser Cavity-Soliton Micro-Combs", Nat. Photonics **13**, no. 6, 384–389 (2019).
26. M.Peccianti, A.Pasquazi, Y.Park, B.E Little, S.T.Chu, D.J Moss, and R.Morandotti, "Demonstration of an ultrafast nonlinear microcavity modelocked laser", Nat. Communications **3,**  765 (2012). doi:10.1038/ncomms1762 (2012).